\documentclass[submission,copyright,creativecommons]{eptcs}
\providecommand{\event}{WCSI 2010} 
\usepackage{breakurl}             

\usepackage{amsmath}
\usepackage{amssymb}
\usepackage{amsfonts}
\usepackage{bm}

\usepackage{multicol}

\usepackage{bcprules}
\usepackage{listings}

\usepackage{graphicx}
\usepackage{subfigure}

\usepackage{amsthm}

\title{Tau Be or not Tau Be? \\ A Perspective on Service Compatibility and Substitutability}
\author{Meriem Ouederni
\institute{University of M\'alaga, Spain}
\email{meriem@lcc.uma.es}
\and
Gwen Sala\"un
\institute{Grenoble INP--INRIA--LIG, France}
\email{Gwen.Salaun@inria.fr}
}
\def\titlerunning{Tau Be or not Tau Be? - A Perspective on Service Compatibility and Substitutability}
\def\authorrunning{M. Ouederni \& G. Sala\"un}
\begin{document}
\maketitle

\begin{abstract}
One of the main open research issues in Service Oriented Computing is
to propose automated techniques to analyse service interfaces. A first
problem, called compatibility, aims at determining whether a set of
services (two in this paper) can be composed together and interact
with each other as expected. Another related problem is to check the
substitutability of one service with another. These problems are
especially difficult when behavioural descriptions ({\it i.e.},
message calls and their ordering) are taken into account in service
interfaces.
Interfaces should capture as faithfully as possible the service
behaviour to make their automated analysis possible while not
exhibiting implementation details. In this position paper, we choose
Labelled Transition Systems to specify the behavioural part of service
interfaces. In particular, we show that internal behaviours ($\tau$
transitions) are necessary in these transition systems in order to
detect subtle errors that may occur when composing a set of services
together. We also show that $\tau$ transitions should be handled
differently in the compatibility and substitutability problem: the
former problem requires to check if the compatibility is preserved
every time a $\tau$ transition is traversed in one interface, whereas
the latter requires a precise analysis of $\tau$ branchings in order
to make the substitution preserve the properties ({\it e.g.}, a
compatibility notion) which were ensured before replacement.
\end{abstract}

\section{Introduction}

The definition of Interface Description Languages (or contract
languages) which provide a good tradeoff between expressiveness and
abstraction level is not a recent research topic. With the advent of
Component-Based Software Engineering in the 90s, many works were
dedicated to the design of such languages, see for
instance~\cite{BeugnardJP99,deAlfaroFSE2001,CanalPT01,PlasilTSE2002}. This
work took a new breath with the recent venue of (Web)
services. Indeed, although the black-box nature of components can be
source of discussion~\cite{BuchiW99,Henriksson07,Puntigam07}, this is
not the case of services since they are deployed and available
on-line, therefore their internal implementation has no reason to be
accessible to users.

Existing IDLs distinguish four interoperability
levels~\cite{BeckerDagstuhl2005}: signature, behaviour (or interaction
protocol), semantics, and quality of service. In this paper, we focus
on the behavioural description level. This level has often been
emphasised as crucial~\cite{PlasilTSE2002,Canal-Poizat-Salaun-08}
because an {\em a-priori} knowledge of every service control flow is
essential to avoid {\em a-posteriori} erroneous executions (such as
deadlock) of a set of interacting services. In the services area,
several notations (Petri nets, transition systems, process algebras,
state diagrams, etc.) have already been proposed to specify the
behavioural part of service interfaces. Here, we have chosen Labelled
Transition Systems which is a simple yet expressive model often used
as semantic foundation to higher-level formalisms. In this model, we
also consider internal or non-observable behaviours using $\tau$
transitions. Internal behaviours are important because analysing
service interfaces may show that they will interact correctly if
observable behaviours only are considered whereas they will actually
behave erroneously due to internal behaviours. These $\tau$
transitions correspond to abstractions of pieces of code ({\it e.g.},
conditions involving variables and functions). This information may be
preserved and may help the designer who wants to compose a set of
services together. However, as far as automatic approaches are
concerned, the analysis of such conditions is difficult because it
deserves to have a full understanding of types and functions used in
those guards.

Once the interface model is defined, several issues have to be worked
out and are still actively studied in the service research community:
service discovery, automatic composition, validation and verification,
adaptation, etc. Here, we focus on two problems (related to one
another) referred to as service compatibility and substitutability (or
replaceability). The first problem aims at checking whether two (or
more) services are {\it compatible}, that is can interact {\it
  properly} until reaching a correct termination state. Several
notions of behavioural compatibility have already been proposed in the
literature. In this paper, we will use three notions for illustration
purposes, namely deadlock-freeness~\cite{CanalPT01},
unidirectional-complementarity and
unspecified-receptions~\cite{YS-ACM97,DZ-ACM83}. The second problem
aims at checking if one service can be {\it substituted} with another
while ensuring that the reconfigured system will behave {\it
  similarly} from a behavioural point of view.

Our goal in this paper is to focus on $\tau$ transitions and first
show that such transitions are necessary in interface models to avoid
erroneous behaviours. Second, we will show that when checking
compatibility and substitutability, these $\tau$ transitions have to be
handled {\it correctly}. One of the main objectives of this paper is
to explain what is meant by {\it correctly} (or {\it properly}, {\it
  similarly}, which are adverbs used before in this introduction). In
particular, their analysis is different from one problem to
another. Compatibility requires to check that observable actions
satisfy the compatibility notion every time an internal behaviour is
traversed in one of the two involved services. On the other hand, the
substitutability verification requires a precise analysis of $\tau$
branchings in order to check that the new service behaves as its
former version. We will illustrate our arguments throughout this paper
with some simple examples.

Our objective is not to present some algorithms and tools to automate
those checks. Such algorithms can be found in related papers, {\it
  e.g.},~\cite{CanalPT01,deAlfaroFSE2001,CernaVZ07}. As far as tool
support is concerned, the compatibility check can be implemented using
Maude~\cite{CDELMMT:2007-book} as presented in~\cite{DOS-Foclasa09},
and the substitutability check can be achieved using
CADP~\cite{CADP2006} and the Bisimulator tool~\cite{RaduTACAS2005}.

The rest of the paper is organized as
follows. Section~\ref{section:model} introduces our behavioural model
of services. Section~\ref{section:comp} first presents some
compatibility notions we use in this paper for illustration purposes,
and then discuss the way $\tau$ transitions should be handled. In
Section~\ref{section:sub}, we tackle the substitutability problem, and
present some solutions to check the substitutability of services with
a special focus on $\tau$ transitions. Finally, we overview existing
works in Section~\ref{section:related} and draw up some conclusions
in Section~\ref{section:conclusion}.

\section{Model of Service Interfaces}
\label{section:model}

We assume that service interfaces are equipped both with a signature
(set of required and provided operations) and a protocol represented
by a {\it Symbolic Transition System} (STS) which is a Labelled
Transition System (LTS) extended with value passing (data parameters
coming with messages). More formally, an STS is a tuple $(A, S, I, F,
T)$ where: $A$ is an alphabet which corresponds to the set of labels
associated to transitions, $S$ is a set of states, $I \in S$ is the
initial state, $F \subseteq S$ is a nonempty set of final states, and
$T  \subseteq S {\scriptsize\setminus} F \times A \times S$ is the transition
relation.
In our model, a {\it label} is either a $\tau$ (internal action) or a
tuple $(m,d,pl)$ where $m$ is the message name, $d$ stands for the
communication direction (either an emission $!$ or a reception $?$),
and $pl$ is either a list of data terms if the label corresponds to an
emission, or a list of variables if the label is a reception.

Notice that, using the STS model, a choice can be represented using
either a state and at least two outgoing transitions labelled with
observable actions (external choice) or branches of $\tau$ transitions
(internal choice). Another possibility would be to keep choice
conditions as part of the model (as done in Symbolic Transition Graph
introduced in~\cite{HennessyL95}), and analyse them using subtyping
relations, see~\cite{ChaeLB08} for instance. However, in the general
case, it is not possible to analyse boolean expressions used in guards
because they can involve variables and functions, and at design-time,
we do not know variable values. Therefore, there is no way to predict
how a choice will behave at run-time. This is why choice or loop
conditions are often made abstract and specified as $\tau$ transitions
in behavioural interfaces.

In our model, no transition can go out from a final state because, in
(Web) services, an implementation explicitly defines a termination
construct ({\it e.g.}, {\sf Terminate} in BPEL), and therefore the
corresponding transition system consists of a transition labelled with
$\tau$ followed by a final state. Such $\tau$ transition can be
minimized if it appears in a sequence ($\tau$-confluence), but this is
not the case if it is involved in a branching structure (a state with
several outgoing transitions).

Synchronizations between services respect a synchronous and binary
communication model. Therefore, two services synchronize if one can
evolve through an emission, the other through a reception, and both
labels have the same message and matching parameters (same number of
parameters with the same type and in the same order).
Internal behaviour cannot be controlled because this corresponds to an
independent evolution of a service, {\it i.e.}, a service can
internally decide to change its state without any apparent or
observable reason.
%
%
The operational semantics of STS is formalised
in~\cite{DOS-Foclasa09}.

This model is simple yet offers a good abstraction level for
describing and analysing service behaviours.  Moreover, STSs can be
easily derived from higher-level description languages such as
Abstract BPEL, see for
instance~\cite{BultanWWW04,SalaunBS06,CamaraMSCOCP09} where such
abstractions were used for verification, composition or adaptation of
Web services.  In the rest of the paper, we will describe service
interfaces only with their corresponding STSs. Signatures can be
deduced from the argument types appearing in STS labels.

\medskip

\begin{figure}[!ht]
\centering
\includegraphics[width=0.6\linewidth,clip]{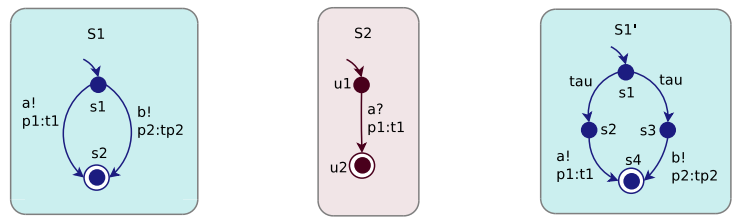}
\caption{{\sf S1} and {\sf S2} interoperate successfully, but {\sf S1'} and {\sf S2} can deadlock}
\label{fig:internal-vs-external}
\end{figure}

{\bf Internal Behaviours.}  Service analysis could be worked out
without taking into account their internal evolution because that
information is not observable from its partners point of view
(black-box assumption). However, keeping an abstract description of
the non-observable behaviours while analysing services helps to find
out possible interoperability issues.  Indeed, although one service
can behave as expected by its partner from an external point of view,
interoperability issues may occur because of unexpected internal
behaviours that services can execute.  For instance,
Figure~\ref{fig:internal-vs-external} shows two versions of one
service protocol without ({\sf S1}) and with ({\sf S1'}) its internal
behaviour. As we can see, {\sf S1} and {\sf S2} can interoperate on
{\sf a} and terminate in final states ({\sf b!} in {\sf S1} has no
counterpart in {\sf S2} and cannot be executed, this is due to
synchronous communication). However, if we consider {\sf S1'}, which
is an abstraction closer to what the service actually does, we see
that this protocol can (choose to) execute a $\tau$ transition at
state {\sf s1} and arrives at state {\sf s3} while {\sf S2} is still
in state {\sf u1}. At this point, both {\sf S1'} and {\sf S2} cannot
exchange messages, and the system deadlocks. This issue would not have
been detected with {\sf S1}.

Now, let us focus on higher-level languages, such as abstract BPEL or
abstract Windows workflow (WF), which are used in the
literature~\cite{MateescuPS08,CuboSCPP08,MartinPimentel-FOCLASA08} as
abstract descriptions (Interface Description Languages) of service
behaviours.  Here we choose WF to illustrate how STSs and in
particular $\tau$ transitions are extracted from this workflow-based
notation.  WF describes service behaviours using a set of basic
activities, {\it e.g.}, {\sf IfElse}, {\sf Listen} and {\sf While},
for which it is useful to keep some $\tau$ transitions in their
respective STS descriptions.

The {\sf IfElse} activity corresponds to an internal choice deciding
which activity has to be performed, {\it e.g.}, sending different
messages using the {\sf WebServiceOutput} activity, depending on the
condition truth value. The corresponding STS contains as many
transitions labelled with $\tau$ as there are branches in the {\sf
IfElse} activity (including the {\it else} branch), see the first
example in Table~\ref{tab:wfvssts}.

Transitions labelled with $\tau$ can describe timeouts, as it is the
case in the {\sf Listen} activity of WF.  This activity waits for
possible receptions ({\sf EventDriven}). If no message is received, a
timeout occurs ({\sf Delay}) which stops the {\sf Listen} activity.
In the STS model, the {\sf Listen} activity is translated into a set
of branches labelled with the receptions used in this activity and a
$\tau$ transition corresponding to the timeout, see the second example
in Table~\ref{tab:wfvssts}.

The {\sf While} activity is used to repeat an activity as long as the
loop condition is satisfied.  Hence, the corresponding STS encodes
this activity using a non-deterministic choice, specified using $\tau$
transitions, between the looping behaviour and the behaviour that can
be executed after the {\sf While} activity (when the condition becomes
false), see the third example in Table~\ref{tab:wfvssts}.

\begin{small}
\begin{table}[!ht]
\centering
\begin{tabular}{|l|c|}
\hline
\multicolumn{1}{|c|}{Abstract WF activity}
& STS description \\

\hline
\hline

\includegraphics[width=0.5\linewidth,clip]{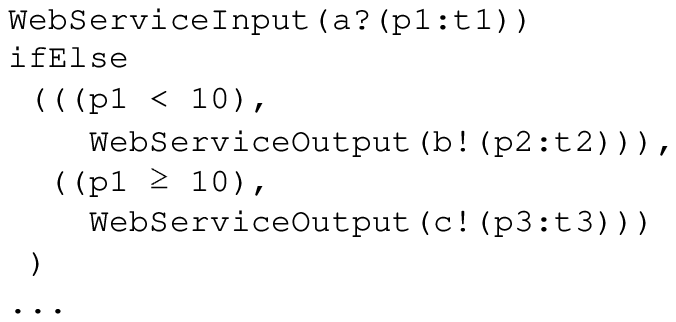}
&
\includegraphics[width=0.25\linewidth,clip]{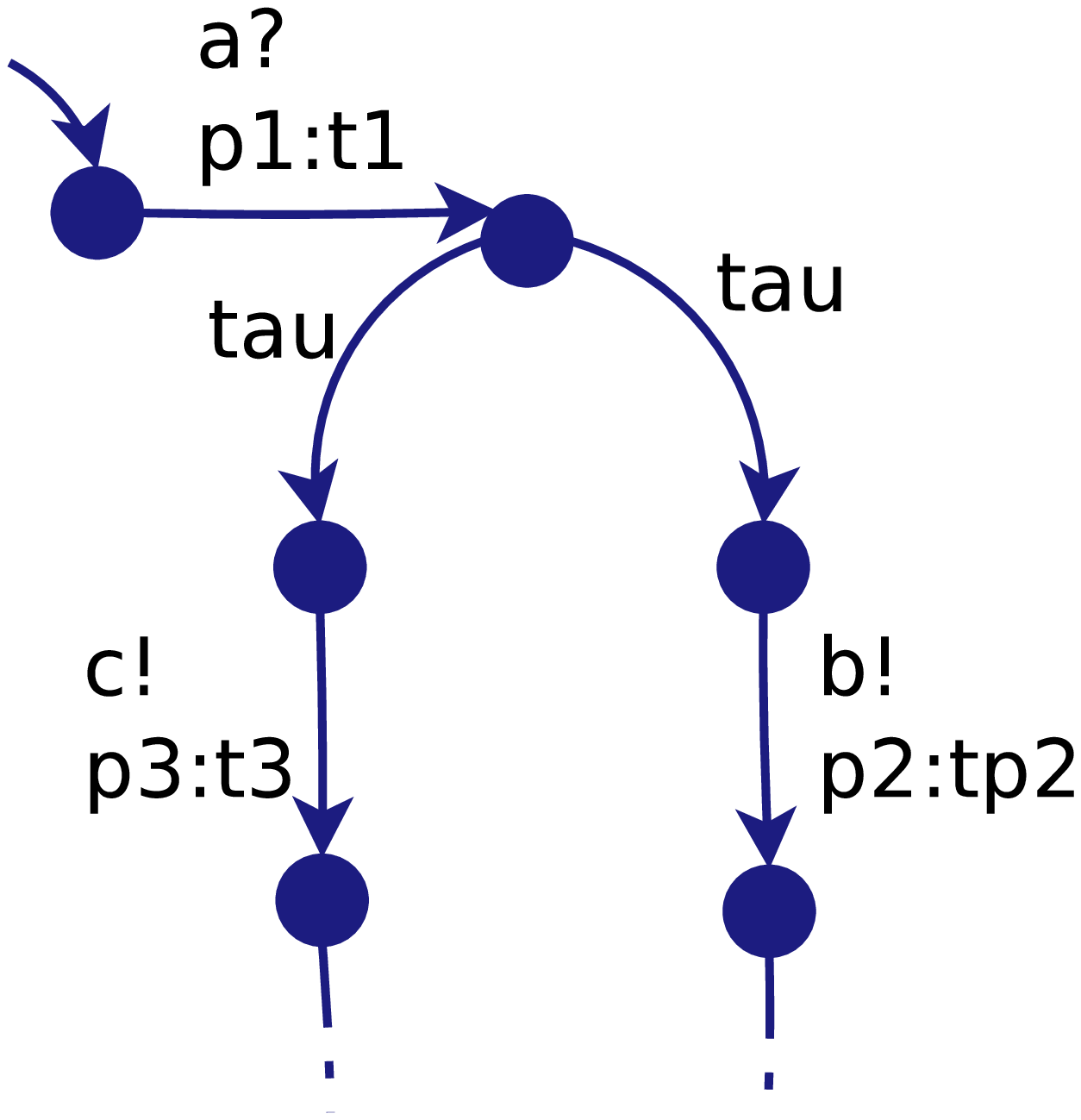} \\

\hline
\hline

\includegraphics[width=0.5\linewidth,clip]{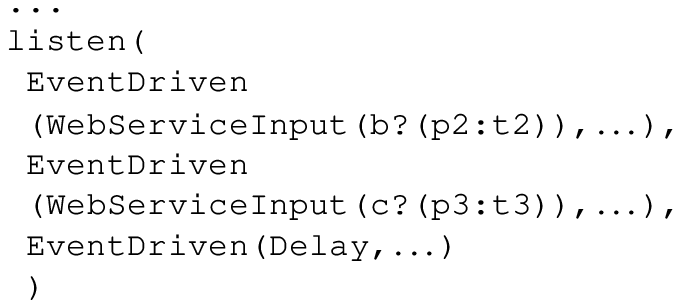}
&
\includegraphics[width=0.25\linewidth,clip]{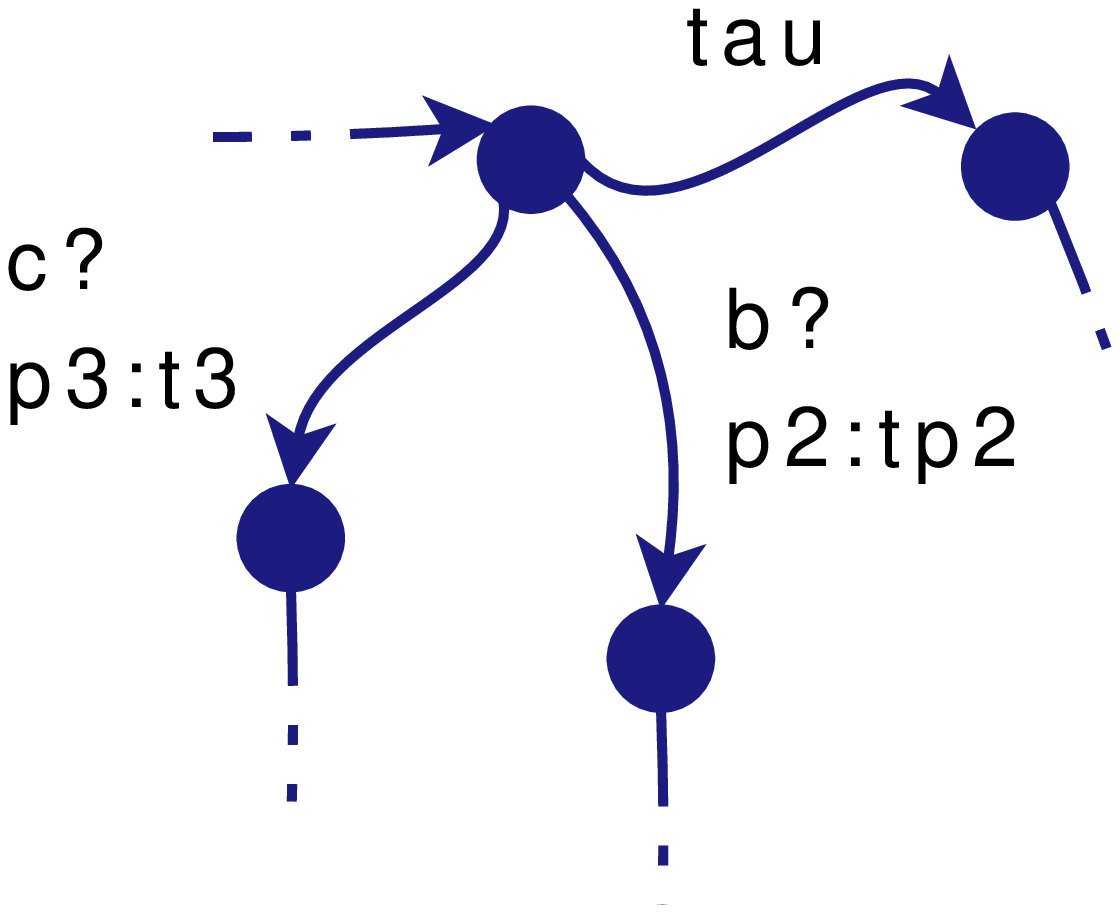} \\

\hline
\hline

\includegraphics[width=0.4\linewidth,clip]{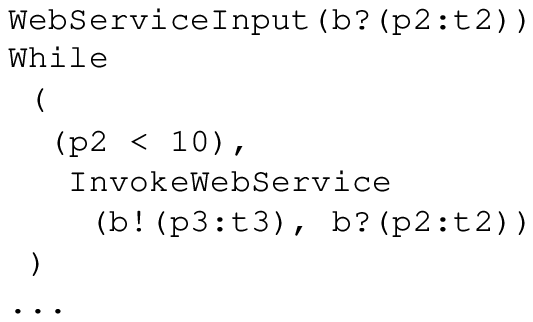}
&
\includegraphics[width=0.25\linewidth,clip]{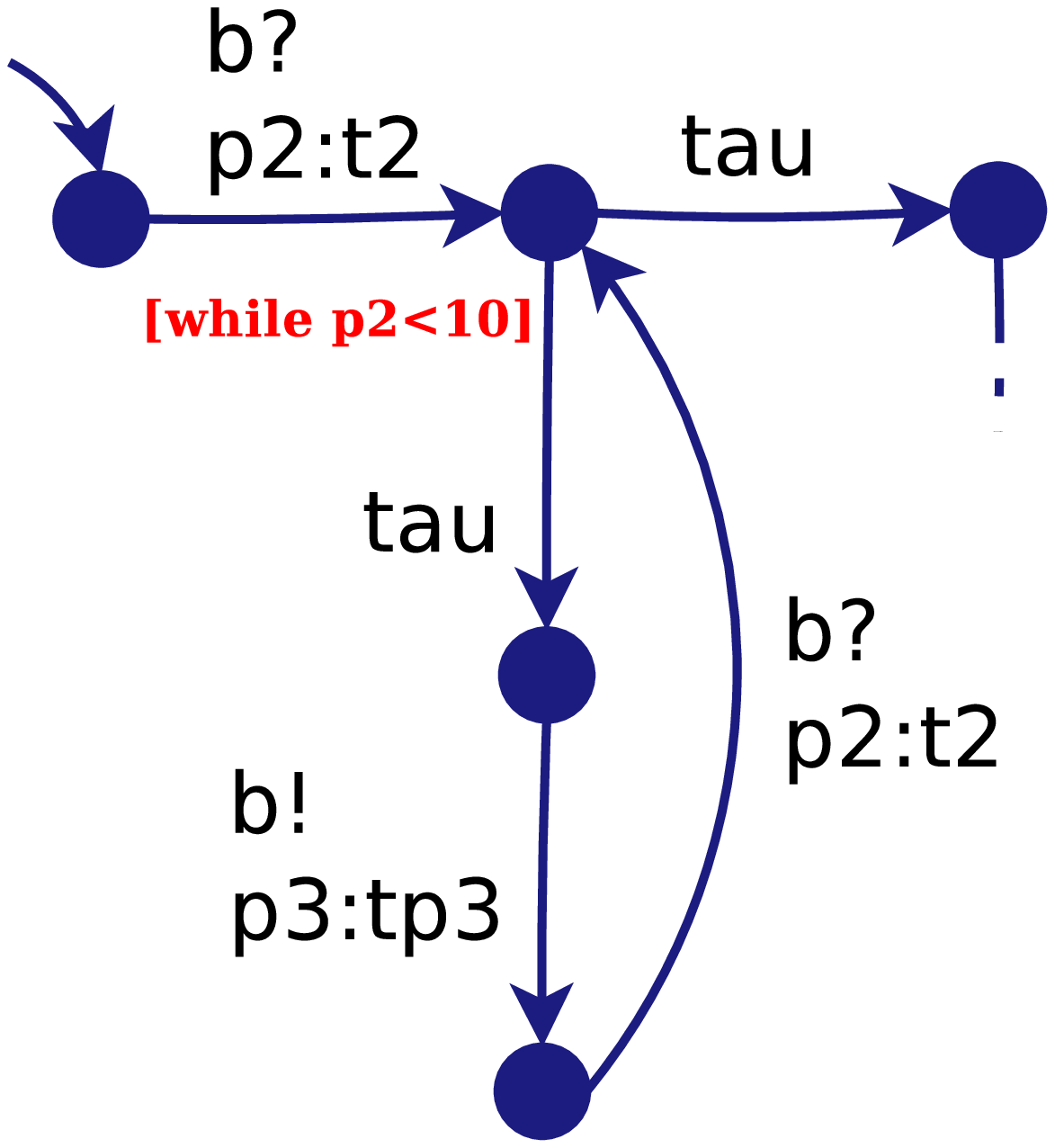} \\
\hline
\end{tabular}
\caption{Examples of abstract WF activities and their corresponding STSs}
\label{tab:wfvssts}
\end{table}
\end{small}

Other abstract WF activities such as {\sf Terminate}, {\sf Parallel}
and {\sf Code} can also generate $\tau$ transitions in the
corresponding STS model.

\section{Compatibility}
\label{section:comp}

In this section, we first present three notions of compatibility,
namely deadlock-freeness, unidirectional-complementarity and
unspecified-receptions. We have chosen these notions because they are
simple to understand, and often used by related work in the
literature~\cite{DZ-ACM83,YS-ACM97,CanalPT01,tes04,DOS-Foclasa09}.  We
will use them in the rest of this paper to illustrate the
discussion. In the second half of this section, we point out the
subtleties of dealing with $\tau$ transitions when checking
behavioural compatibility.

\subsection{Compatibility Notions}
\label{section:compnotions}

\begin{figure}
\centering
\includegraphics[width=0.6\linewidth,clip]{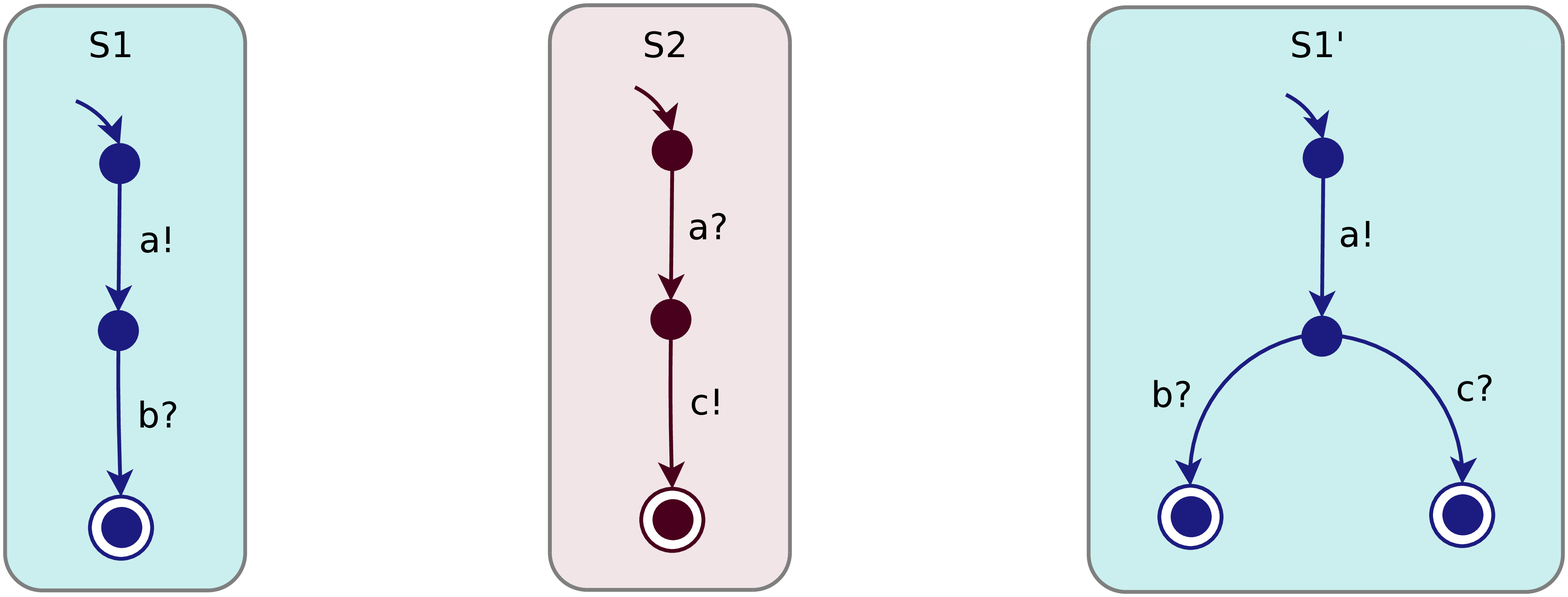}
\caption{Deadlock-freeness compatibility}
\label{fig:dead}
\end{figure}

\noindent\textbf{Deadlock-freeness.} This notion says that two service
protocols are compatible if and only if, starting from their initial
states, they can evolve together until reaching final states.
Figure~\ref{fig:dead} presents a simple example to illustrate this
notion. {\sf S1} and {\sf S2} are not compatible because after
interacting on action {\sf a}, both services are stuck. On the other
hand, {\sf S1'} and {\sf S2} are deadlock-free compatible since they
can interact successively on {\sf a} and {\sf c}, and then both
terminate into a final state.

\medskip

\begin{figure}
\centering
\includegraphics[width=0.6\linewidth,clip]{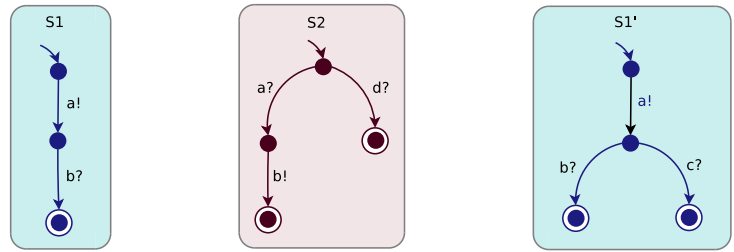}
\caption{Unidirectional-complementarity compatibility}
\label{fig:sim}
\end{figure}

\noindent \textbf{Unidirectional-complementarity.}  Two services are
compatible with respect to this notion if and only if there is one
service which is able to receive (send, respectively) all messages
that its partner expects to send (receive, respectively) at all
reachable states. Hence, the ``bigger'' service may send and receive
more messages than the ``smaller'' one. Additionally, both services
must be free of deadlocks.
This notion is different to what is usually called simulation or
preorder relation~\cite{Milner89} because the two protocols under
analysis here aim at being composed, and accordingly present opposite
directions. However, both definitions share the inclusion concept: one
of the two protocols is supposed to accept all the actions that the
other can do.
Figure~\ref{fig:sim} first shows two services {\sf S1} and {\sf S2}
which respect this unidirectional-complementarity compatibility: all
actions possible in {\sf S1} can be captured by {\sf S2}. However,
{\sf S2} does not complement {\sf S1'} because {\sf S2} is not able to
synchronize on action {\sf c} with {\sf S1'}.

\medskip

\begin{figure}
\centering
\includegraphics[width=0.6\linewidth,clip]{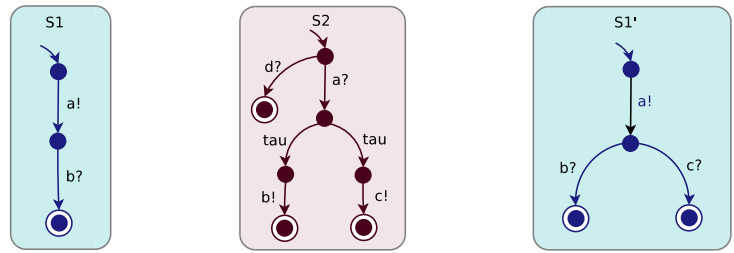}
\caption{Unspecified-receptions compatibility}
\label{fig:unspecified}
\end{figure}

\noindent \textbf{Unspecified-receptions.}  This definition requires
that if one service can send a message at a reachable state, then the
other service must receive that emission. Furthermore, one service is
able to receive messages that cannot be sent by the other service,
{\it i.e.}, there might be additional unmatched receptions.  It is
also possible that one protocol holds an emission that will not be
received by its partner as long as the state from which this emission
goes out is unreachable when protocols interact together.
Additionally, both services must be free of deadlocks.
In Figure~\ref{fig:unspecified}, {\sf S1} and {\sf S2} are not
compatible because {\sf S1} cannot receive all actions that {\sf S2}
can send ({\sf c!}). But {\sf S1'} and {\sf S2} are compatible because
all emissions on both sides have a matching reception on the other.


The reader interested in the formal definitions for these
compatibility notions can refer to~\cite{tes04,DOS-Foclasa09}.

\subsection{Internal Behaviours}
\label{section:tau-handling}


Compatibility checking verifies that two interacting services fulfill
each other's requirements.  Interaction between services basically
depends on synchronisations over observable actions and then can be
defined using a criterion set on these observable actions (see for
instance the compatibility notions we presented previously).
Since services can evolve independently through some non-controllable
$\tau$ transitions, the behavioural compatibility requires that each
internal evolution must lead both services into a state in which the
criterion is satisfied. This means that every time a $\tau$ transition
is traversed in one of the two STSs, the compatibility must be checked
again on the target state. This way to process $\tau$ transitions
leads to a unique way to handle them all along the compatibility
checking.
This is not the case in the context of service substitution where
services can be compared according to different ways of dealing with
their internal behaviours, similarly to what is achieved in
equivalence checking~\cite{Milner89,BHR84,GlabbeekW96} (see
Section~\ref{section:sub} for more details on the substitutability
problem).

\begin{figure}
\centering
\includegraphics[width=0.4\linewidth,clip]{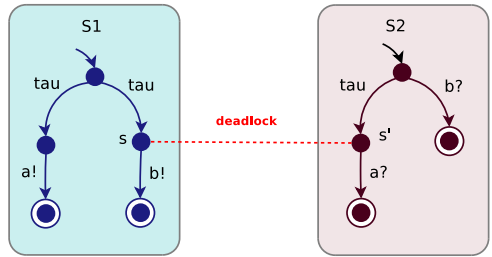}
\caption{$\tau$ transitions need to be analysed}
\label{fig:deadtau}
\end{figure}

Let us illustrate these ideas on a couple of examples. First of all,
Figure~\ref{fig:deadtau} shows that it is not enough to focus on
observable actions when checking service compatibility: $\tau$
transitions must be analysed as well.  In this example, both services
can interact on {\sf a} and {\sf b} from an observational point of
view, {\it i.e.}, considering only observable traces without $\tau$
transitions. However, if the compatibility check does not analyse only
observational actions but also internal ones, a deadlock is
detected when services {\sf S1} and {\sf S2} move to state {\sf s} and
{\sf s'}, respectively, by executing a $\tau$ transition. Therefore,
these two services are not deadlock-free compatible.

\begin{figure}
\centering
\includegraphics[width=0.4\linewidth,clip]{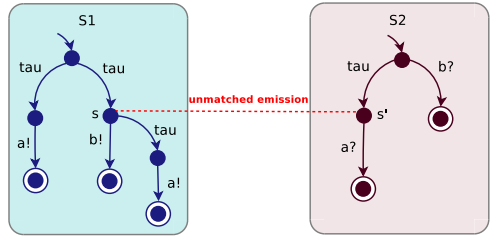}
\caption{$\tau$ transitions matching is not necessary}
\label{fig:urtau}
\end{figure}

The question now is: how $\tau$ transitions are supposed to be
analysed when checking compatibility? Similarly to equivalence
checking, one may want to match $\tau$ transitions appearing in both
service interfaces together. As an example, observational (or weak)
equivalence~\cite{Milner89} checks that one $\tau$ on one side matches
with a sequence of zero or more $\tau$ on the other. Figure~\ref{fig:urtau} shows
an example in which the state matching respects this weak
relation\footnote{Actually, services {\sf S1} and {\sf S2} are
  equivalent {\it wrt.} the observational relation if directions in
  one service are reversed as follows: $\overline{l!}=l?$,
  $\overline{l?}=l!$. In our model, messages may come with parameters,
  and this check would also require to remove parameters
  beforehand.}. Nevertheless, these services are not compatible with
respect to the unspecified-receptions compatibility\footnote{Here, we
  chose a special example where no additional receptions appear in
  both services, and the parallel with equivalence checking is
  therefore easier to make.}, because {\sf S1} can evolve to state
{\sf s} by executing a $\tau$ transition and {\sf S2} to {\sf s'}, and
in this configuration, action {\sf b!} in {\sf S1} has no counterpart
in service {\sf S2} in state {\sf s'}.

To sum up, in order to check compatibility, $\tau$ transitions need to
be analysed, and one has to check after each $\tau$ transition that
the compatibility notion is verified by the forthcoming observational
actions. We also claim that reasoning on the $\tau$ branchings as done
in equivalence checking (matching $\tau$ appearing in both interfaces)
is not useful when one checks compatibility. Indeed, in order to
ensure correct interactions, we do not want to match one service
internal actions with those of its partner. This would be meaningless
in a composition situation because these actions are non-controllable
from a partner point of view, and do not have anything to see with one
another. We only need to check that their observable actions behave as
defined in the compatibility notion in spite of possible $\tau$
transitions.

\section{Substitutability}
\label{section:sub}

Substitutability aims at replacing one service embedded as part of a
larger system with another service such that the entire system is able
to interact as before ({\it i.e.}, respecting the same compatibility
notion). In this paper, we focus on context-dependent approaches, {\it
  i.e.}, where partners (sometimes called environment) are defined and
known.

\begin{figure}
\centering
\includegraphics[width=0.6\linewidth,clip]{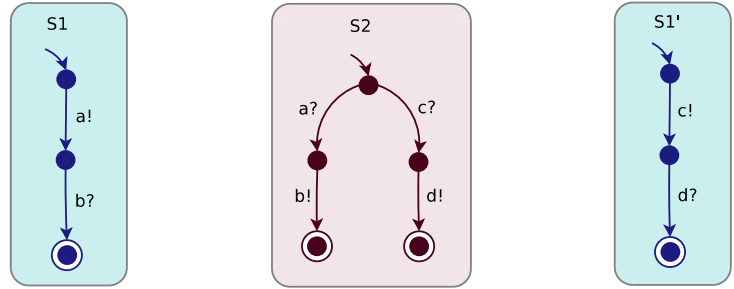}
\caption{Checking substitutability as compatibility may be misleading}
\label{fig:formulations}
\end{figure}

First of all, the substitutability problem has two formulations in the
literature. Suppose we have a system consisting of two services {\sf
  S1} and {\sf S2}, and both services are compatible {\it wrt.} a
compatibility notion {\sf C}. Imagine now that we want to substitute
service {\sf S1} with a new service {\sf S1'}. A first way to check if
this substitution is possible is to verify that {\sf S1'} and {\sf S2}
are compatible {\it wrt.} compatibility {\sf C}. A second way is to
compare {\sf S1} and {\sf S1'} to ensure they are related by a certain
{\it relation}. Both solutions are valid, however the first
formulation can be misleading and for this reason, we will focus on
the second in the rest of this section. To illustrate this point, see
the example given in Figure~\ref{fig:formulations} where we consider
for example the unspecified-receptions compatibility. {\sf S1} and {\sf
  S2} are compatible {\it wrt.} this notion. As far as the first
formulation above is considered, {\sf S1'} can substitute {\sf S1}
because {\sf S1'} is compatible with {\sf S2} (no reachable emissions
without counterpart in both protocols). Nevertheless, {\sf S1} and
{\sf S1'} have completely different behaviours and therefore fulfill
different objectives as well (imagine for instance that action {\sf a}
corresponds to a search in a database, and action {\sf c} to a
modification of that database).

\begin{figure}
\centering
\includegraphics[width=0.6\linewidth,clip]{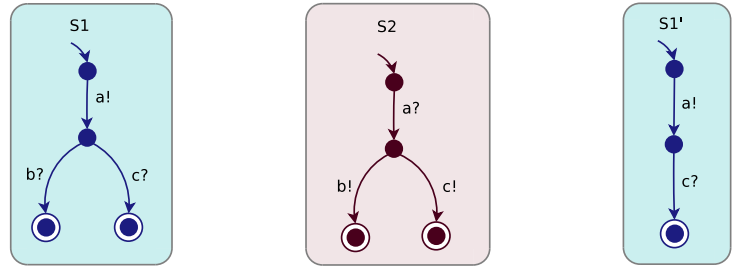}
\caption{Simulation is too weak to check substitutability}
\label{fig:simulation}
\end{figure}

In our model, we chose a level of abstraction where we replace guards
with $\tau$ transitions. An alternative approach is to keep guards and
use subtyping techniques, see for instance~\cite{ChaeLB08}, for
analysis purposes when checking the substitutability problem. Since our
model considers $\tau$ transitions, we can use strong notions such as
equivalences~\cite{Milner89} (or bisimulations), or more flexible ones
such as simulation~\cite{Milner89} or behavioural
subtyping~\cite{LiskovW94}\footnote{Refinement is also a notion used
  for the substitution problem, see~\cite{BeyerCH05,Hameurlain07} for
  example. This notion is stronger than
  subtyping~\cite{Hameurlain07}, but we will not talk about it in
  this paper because our goal is to focus on $\tau$ handling and not
  to give a survey on substitution notions.}. In the general case (for
any compatibility definition as one of those presented in
Section~\ref{section:comp}), simulation or subtyping can be too loose.
Let us focus first on the simulation (or preorder)
notion. Intuitively, all the actions possible in one transition system
must appear in the other. In such a case, we say that the ``bigger''
protocol simulates the ``smaller''. In Figure~\ref{fig:simulation}, we
show that this notion is too weak to always preserve
compatibility. This example shows two services {\sf S1} and {\sf S2}
which are unspecified-receptions compatible. However, if we replace
{\sf S1} with {\sf S1'} where {\sf S1} simulates {\sf S1'}, then there
is an emission ({\sf b!}) in {\sf S2} which has no counterpart in {\sf
  S1'}, therefore {\sf S1'} and {\sf S2} are not compatible.

\begin{figure}
\centering
\includegraphics[width=0.6\linewidth,clip]{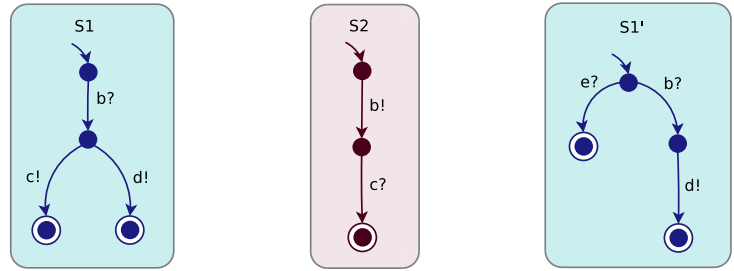}
\caption{Behavioural subtyping is too weak in the general case}
\label{fig:bsubtyping1}
\end{figure}

\begin{figure}
\centering
\includegraphics[width=0.6\linewidth,clip]{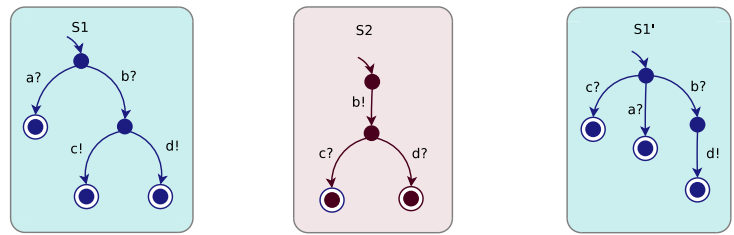}
\caption{Behavioural subtyping works for precise compatibility notions, {\it e.g.}, unspecified-receptions}
\label{fig:bsubtyping2}
\end{figure}

Regarding behavioural subtyping, different definitions exist in the
literature for the substitutability problem, see for
instance~\cite{CanalPT01,BeyerCH05,Hameurlain07}. We illustrate here
with the definition proposed in~\cite{BeyerCH05} where {\it ``the
  algorithm for substitutivity checking verifies that service A
  demands fewer and fulfills more constraints than service B''}. In
terms of transition systems, this means that a service can replace
another if it can have more receptions and less emissions. Again, in
the general case, this definition is not strong
enough. Figure~\ref{fig:bsubtyping1} gives a simple example where two
services {\sf S1} and {\sf S2} are deadlock-free compatible, but {\sf
  S1'} and {\sf S2} are not, even if {\sf S1'} is a behavioural
subtype of {\sf S1} according to the definition quoted above. Let us
emphasise here that our claim focuses on the general case (any
compatibility notion). If we consider a precise compatibility notion,
it can be demonstrated that this behavioural subtyping relation is
enough. This is the case for instance with the unspecified-receptions
compatibility (see Figure~\ref{fig:bsubtyping2} for an example)
because the new service can have more receptions and less
emissions. As a consequence, all the emissions in the service which
does not change are still captured (the new service preserves all its
former receptions and may have more). Moreover, the new service can
only have less emissions compared to its former version, and since all
the emissions in the old service had a counterpart in its partner, the
new service will have corresponding receptions as well.

Equivalences are strong yet suitable relations to check the
substitutability problem, because they preserve all observable actions
and then the compatibility notion should be preserved as
well. However, different equivalence relations exist, and they handle
differently internal behaviours. In this paper, we will focus on some
well-known equivalence relations, namely strong, branching, weak, and 
trace equivalence, from the strongest to the weakest notions
(see~\cite{ChapterHPA-Intro} for more details on these notions and
their formal relationship).

As far as substitutability is concerned, a strong equivalence or
bisimulation~\cite{Milner89} is too strong because it requires to
match not only observable actions but also $\tau$
transitions. Perfectly matching these internal transitions does not
make sense in the Web services area because two service
implementations can sligthly differ yet exhibit exactly the same
behaviour from an external point of view.

\begin{figure}
\centering
\includegraphics[width=0.6\linewidth,clip]{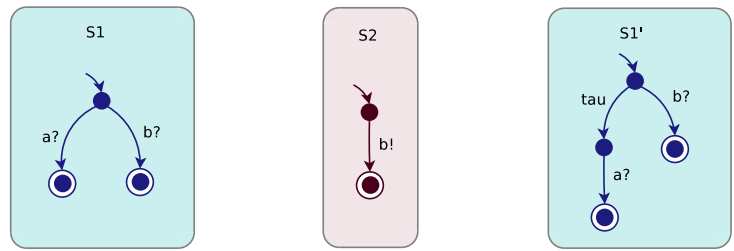}
\caption{Trace equivalence is too weak to check substitutability, and does not preserve compatibility}
\label{fig:trace}
\end{figure}

At the other extremity, trace equivalence is too weak because this
relation only analyses the observable behaviour, and does not preserve
compatibility. Figure~\ref{fig:trace} shows an example where {\sf S1}
and {\sf S2} are two services respecting a deadlock-freeness
compatibility. However, if we replace {\sf S1} by {\sf S1'}, even if
{\sf S1} and {\sf S1'} are trace equivalent, {\sf S1'} and {\sf S2}
are not deadlock-free compatible because a deadlock occurs if {\sf
  S1'} decides to execute the $\tau$ transition.

Weak and branching equivalences are the strongest of the weak
equivalences~\cite{ChapterHPA-Intro}. These two notions preserve
behavioural properties (does not add deadlocks for instance) on
observable actions. Consequently, these two equivalence relations are
adequate to verify the substitutability of one service by
another. Such results were formally proven for weak
equivalence~\cite{CernaVZ07}.  Branching equivalence is stronger than
weak equivalence, and is checked more efficiently from a computational
point of view, so it should be preferred if huge protocols are
involved when checking substitutability.

%

\section{Related Work}
\label{section:related}
In this section, we overview works existing on the compatibility and
substitutability questions, with a specific focus on approaches
handling internal behaviours in their model and solutions.

\subsection{Compatibility}

To the best of our knowledge, here are the
approaches~\cite{CanalPT01,Hameurlain05,wu2009computing,deAlfaroFSE2001,DOS-Foclasa09}
which take $\tau$ transitions into account in their description
models.
In~\cite{Hameurlain05}, the author relies on a bisimulation algorithm
to define the compatibility of (Web) services which are described
using Petri nets.  The bisimulation-based compatibility associates
$\tau$ transitions in the same way as Milner's strong
equivalence. Matching $\tau$ transitions as done in the strong equivalence
does not make sense when checking services compatibility in our
opinion.

In~\cite{CanalPT01}, a compatibility notion based on the
$\pi$-calculus considers two services to be compatible if they are
deadlock-free. In~\cite{deAlfaroFSE2001}, the authors rely on an
automata-based model and define context-dependent compatibility. This
work considers two interfaces to be compatible if their product can be
composed with a third component and this composition is
deadlock-free. In these two works, the authors propose to analyse
$\tau$ transitions similarly to what is introduced in
Section~\ref{section:tau-handling}, {\it i.e.}, each internal
evolution must lead the system into states where the deadlock-freeness
is preserved.

A $\pi$-calculus description model is also used
in~\cite{wu2009computing} where two services are compatible if there
is always at least one sequence of interactions that make them reach
final states. This notion is quite weak when composing services
because the deadlock-freeness property cannot be
guaranteed. In~{\cite{wu2009computing}, $\tau$ transitions only appear
  as the visible result of synchronisations (as defined in the
  $\pi$-calculus or CCS semantics). Then, two services are considered
  compatible if their composition can engage a sequence of $\tau$
  actions until reaching final states. Therefore, no particular
  processing is associated to $\tau$ transitions in this approach.

In a previous paper~\cite{DOS-Foclasa09}, we considered an
automata-based model and proposed a generic framework which
automatically checks service protocols according to a compatibility
notion passed as parameter. Three strategies for handling $\tau$
transitions were implemented, namely strong, weak and trace. These
strategies are inspired from the ways of associating internal
behaviours proposed by the strong, weak and trace equivalence
relations. Considering different ways of dealing with $\tau$
transitions does not impact the result of the compatibility check, but
adds an additional analysis on $\tau$ branchings.

\subsection{Substitutability}

Here again, our goal is not to survey all existing works but only
those which use internal behaviours in their interface model. First,
Hameurlain~\cite{Hameurlain05} addresses the substitutability of
component protocols described with Petri nets. The substitutability
notion used in this paper is a strong bisimulation as introduced by
Milner in~\cite{Milner89}.  Replacing components using this relation
enables to preserve system compatibility. However, it is a very strict
relation as far as the matching of $\tau$ transitions is concerned,
and weaker relations may be enough to preserve this compatibility.

In~\cite{CernaVZ07}, the authors check component substitutability
using weak bisimulation.  They show that whenever there is a system in
which a component is replaced with an observationally equivalent one,
the system remains equivalent to the former one. This relation is less
restrictive than strong bisimulation used in~\cite{Hameurlain05}.

More recently, ~\cite{ChaeLB08} used a Finite State Machine (FSM)
model to formalise a substitutability notion for Web services which
preserves compatibility. The authors consider a symmetric approach
which requires that services must have the same traces. In this paper,
pre/post-conditions are used rather than $\tau$
transitions. Therefore, the authors compare these conditions using a
subtyping relation: the pre-conditions of an old service must be
simulated by those of the new service and the post-conditions of the
new service must be simulated by those of the old service.

\section{Concluding Remarks}
\label{section:conclusion}

In this paper, we have focused on behavioural models of service
interfaces, especially those involving internal behaviours. Those
behaviours are essential because if they are not taken into account in
service models, the composition or substitution of services may cause
erroneous executions. We have discussed various solutions to handle
internal behaviours when checking compatibility and substitutability
of services. Our conclusions are the following: (i)~when checking
compatibility, the notion to be ensured has to be verified after every
internal behaviour appearing in each behavioural interface, and
(ii)~when checking substitutability, behavioural models need to be
equivalent {\it wrt.} a relation stronger enough (such as weak or
branching equivalence) to preserve all properties on observable
behaviours.


Now, we would like to conclude with four challenges which are still
some open issues in the context of the compatibility and
substitutability checking: (i)~generalising existing approaches to
consider not only two services but a set of services (compatibility of
$n$ services, substitution of $k$ services involved in a system by $m$
new services, etc.), (ii)~considering an asynchronous communication
model ({\it e.g.}, based on message queues), (iii)~proving that
branching equivalence is better than weak equivalence when checking
service substitutability, and (iv)~not only returning a boolean result
but, if services cannot be properly composed or replaced ({\it false}
result), detecting the mismatches and measuring the
compatibility/substitutability degree separating both protocols.

\medskip
\noindent\textbf{Acknowledgements.} The authors thank Radu Mateescu
for interesting comments on an earlier version of this paper, and
Francisco Dur\'an for fruitful discussions on this topic.  This work
has been partially supported by the project TIN2008-05932 funded by
the Spanish Ministry of Innovation and Science (MICINN) and FEDER.

\bibliographystyle{eptcs}
\bibliography{biblio}

\end{document}